# Network structure of subway passenger flows

**Q Xu[1], B H Mao[1, 2] and Y Bai[1]**

[1] MOE Key Laboratory for Urban Transportation Complex Systems Theory and Technology, Beijing Jiaotong University, Beijing, 10044, China
[2] Integrated Transportation Research Center of China, Beijing Jiaotong University, Beijing, 10044, China
E-mail: xuqi2015@bjtu.edu.cn

**Abstract**. The results of transportation infrastructure network analyses have been used to analyze complex networks in a topological context. However, most modeling approaches, including those based on complex network theory, do not fully account for real-life traffic patterns and may provide an incomplete view of network functions. This study utilizes trip data obtained from the Beijing Subway System to characterize individual passenger movement patterns. A directed weighted passenger flow network was constructed from the subway infrastructure network topology by incorporating trip data. The passenger flow networks exhibit several properties that can be characterized by power-law distributions based on flow size, and log-logistic distributions based on the fraction of boarding and departing passengers. The study also characterizes the temporal patterns of in-transit and waiting passengers and provides a hierarchical clustering structure for passenger flows. This hierarchical flow organization varies in the spatial domain. Ten cluster groups were identified, indicating a hierarchical urban polycentric structure composed of large concentrated flows at urban activity centers. These empirical findings provide insights regarding urban human mobility patterns within a large subway network.

## 1. Introduction

Recent transportation network studies have received considerable attention from the physics community. Various types of network infrastructure graphs have been studied, including road [1-2], railway [3-4], airline [5-6], and public transportation networks [7-8]. Subway transport is an important public transportation sector with a variety of fascinating network properties. Previous studies have provided valuable insights into the structure and dynamics of subway infrastructure networks, including node centrality [9], scale-free patterns [10-11], small-world properties [12-13], network efficiency [14-16], network vulnerabilities [17-18] and network growth patterns [19]. These studies typically focused on the complex network topology of subway systems and not network traffic flows due to data collection challenges.

However, the subway system network analyses that are based on physical infrastructure configurations are significantly limited, as they do not account for real-life traffic patterns in the network [20-22]. In addition, these limitations may provide an incomplete view of the network functions because passenger transport is the ultimate goal of any transportation system. For example, the betweenness centrality metric has been used to identify key nodes in subway systems based on the assumption that each pair of nodes exchanges the same amount of traffic [23]. However, subway network passenger flows are exhibit size and spatiotemporal heterogeneities. Therefore, the most important topological nodes or links may not necessarily transport the most traffic [24], and geographical patterns formed by spatial flow distributions and betweenness may substantially differ in transportation networks [20].

The use of GPS units and mobile devices to track human mobility patterns has received growing interest from the field of urban planning [25-26]. Some subway network traffic flows have been conducted, including spatiotemporal human mobility pattern studies in the London Underground [27] and spatiotemporal density and train trajectory analyses of the Singapore Mass Rail Transit [28]. However, these studies do not account for the underlying physical topology of subway systems. A comprehensive system view is required to analyze physical network structures, traffic patterns and network functions in subway systems.

Few passenger flow analyses have considered the underlying physical topology of subway systems. Lee et al. [29] studied passenger flows in the Metropolitan Seoul Subway by building a weighted flow network and determining the network flow distributions. However, the implications of the results were not interpreted. Lee et al. also proposed a master equation approach to further study subway system passenger flows and deduced an evolution equation for passenger distributions. The equation encompasses passenger flow distributions in the Seoul Subway system as well as the growth of Seoul [30]. Soh et al. [24] analyzed passenger flows from the Singapore Mass Rail Transit using a complex weighted network perspective. They found that the network is almost completely topologically connected, but dynamic flow-based measures highlighted hub nodes that experience large passenger flows. Roth et al. [31] utilized trip data from London Underground passengers to study passenger flow patterns and properties to determine the structure and organization of the city. These existing studies have used weighted network-based approaches to determine passenger flow patterns taking trips between adjacent stations. However, the passenger flow patterns that consider both all passengers taking trips between two adjacent stations and using one station (i.e., entering and leaving one station) have not yet been studied.

This paper provides a general framework for developing public transportation passenger movement network models within a city. This study begins by briefly presenting the activity chain of a typical subway trip. We then utilize the trip data obtained from the Beijing Subway System and explicitly include two data sets, including passengers flowing along tunnels (links) and moving in stations (nodes). Next, we contribute a complex directed weighted network analysis to explore the network characteristics of individual movement patterns in terms of flow size distributions, temporal patterns and hierarchical clustering structures. Our findings suggest that public transportation passenger flow networks exhibit rich structures that can be analyzed via complex systems approaches.

This paper is organized as follows. Section 2 presents the activity chain of a typical subway trip and describes the data set. Section 3 constructs the passenger flow networks that form the subway network topologies by incorporating trip data. The statistical properties of subway passenger movement patterns are then presented in Section 4.

## 2. Passenger flows and variables

### *2.1. Passenger flow process*

This study analyzes subway passenger flows using a network with $N$ nodes connected by $L$ bi-directional links. Figure 1 illustrates the typical activities of individual passengers during a subway trip. First, a passenger semi-continuously arrives at an origin station according to a time-dependent pattern during a given period. They then scan their ticket to access the train services that are scheduled based on discrete time stamps. During the same period, passengers arriving at their origin stations make their way to the platform and wait for or board the latest train. They eventually reach their destination and leave the services area by having scanning their ticket at the final destination.

The individual passenger movements within a subway network will result the passengers flowing along links and passengers entering and leaving stations, which involve passenger flows through a network and exogenous flows within a system. Therefore, these individual passenger movements within a subway system can be represented by a hybrid process, where passengers enter/leave their origin/destination stations or nodes, and the resulting flows along tunnels or links are based on train services.

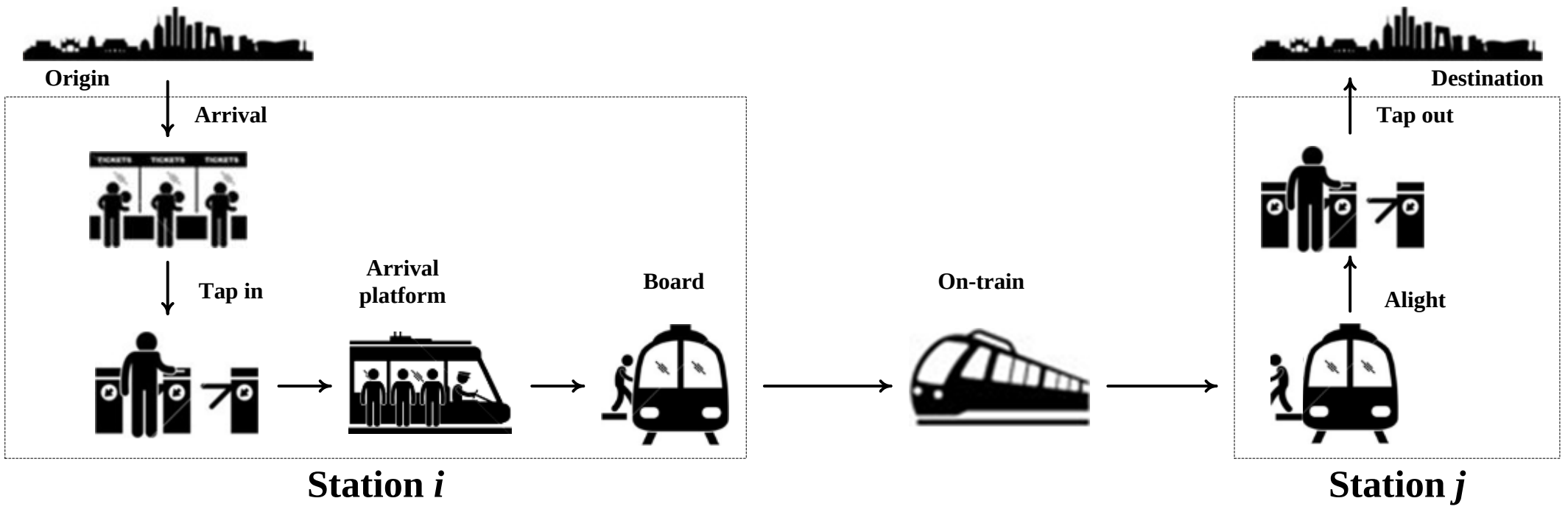

Figure 1 Activity chain of a typical subway trip

*2.2. Description of data*

We consider the Beijing Subway System (BSS), which serves as a major transportation mode in the urban and suburban districts of the Beijing municipality, China. As of May 2015, the network includes 18 lines, 319 unique stations ($N$) and 612 tunnels between stations ($L$). This study analyzes passenger flows in a subsystem operated by the Beijing Mass Transit Railway Operation Corporation. The subnetwork consists of 14 lines, 203 statins and 454 tunnels. The passenger flow analysis is based on a single day passenger trip data that was collected using smart Yikatong card (an electronic ticketing system used to record the passenger movements within the network) transaction data from April 15, 2014. The dataset encompasses 6,161,646 individual passengers on the specific day. The dataset includes public transportation passenger movements within the network for each trip, including the origin and destination stations as well as the corresponding trip travel time.

A passenger flow matrix $\boldsymbol{A}$ was constructed to analyze the BSS network passenger trip data. The elements of $\boldsymbol{A}$ represent the number of passengers taking trips between a pair of adjacent stations over a given period time. Flow matrices $\boldsymbol{B}$ and $\boldsymbol{C}$ were also created. These matrices represent the number of passengers arriving at and leaving one station over a given period time. Note that we divide the data set into half-hour pieces. Therefore, a period of $\Delta t$=30 mins is the time interval used to calculate the aggregated number of individual person movements. These flow matrices can be analyzed over several time intervals within a typical day to explicitly define movement patterns within the network.

## 3. Passenger flow network

*3.1. Notation*

A passenger flow network is represented as a directed weighted graph $G$ with $N$ nodes and $L$ links, and an associated weighted adjacency matrix $\boldsymbol{F}=\{f_{i\to j}\}$, which represents the flow from station $i$ to station $j$:

$$f_{i\to j} = \text{flow from } i \text{ to } j \tag{1}$$

The subscripts $i$ and $j$ represent the destination and source stations, e.g., $f_{i\to j}$ denotes the departure flow (outflow) of station $i$ moving in the $i\to j$ direction, while $f_{j\to \mathrm{i}}$ denotes the arrival flow (inflow) of station $i$ moving in the $j\to i$ direction. $f_{i\to\cdot}$ denotes the total outflow of station $i$, while $f_{\cdot\to i}$ denotes the total inflow of station $i$. The subscripts in this study refer to a pair of adjacent stations unless otherwise noted.

In addition to flows between a pair of stations, a passenger flow network includes a flow moving into station $i$ from outside the network and a flow moving out of the network from the same station. As presented in Section 2, the passenger flows from outside the network to node $i$ correspond to all passengers entering a station from nearby areas or via other transportation modes. The flows from node $i$ to outside the network correspond to all passengers leaving the station to end a trip. Let the flow into node $i$ from outside the network be denoted by $v_{0\to i}$, and the flow outside of the network from node $i$ be denoted by $u_{i\to 0}$:

$$v_{0\to i} = \text{flow from outside to } i \tag{2}$$

$$u_{i\to 0} = \text{flow from } i \text{ to outside} \tag{3}$$

where the surrounding environment that interacts with the network is labeled compartment 0 (zero).

Subway system flow is conserved at all nodes based on unconstrained facility capacities. However, major subways suffer from crowding and traffic congestion. Accordingly, crowding critically affects passenger behaviors, disrupts train operations, inconveniences passengers and creates delays. Therefore, inflow does not always equal outflow. Based on [32], we refer to the flow into and out of node $i$ as the *throughflow*, represented by $T_i^{in}$ and $T_i^{out}$:

$$T_i^{in} \equiv v_{0\to i} + \sum_j f_{j\to i} \tag{4}$$

$$T_i^{out} \equiv u_{i\to 0} + \sum_j f_{i\to j} \tag{5}$$

*3.2. Topology*

A directed weighted passenger flow network is created from a subway network infrastructure topology by incorporating trip data. Figure 2 presents a simplified passenger flow network example using $N$=8 nodes and $L$=14 directed links. We first represent the subway network infrastructure topology based on the L-space representation [7, 14, 20-23], which depicts the original configuration of real-life transportation networks, as shown in figure 2(a). In L-space, subway stations are nodes, and two stations are only connected if they are directly physically connected. However, pairs of adjacent stations are often joined by double-track tunnels, where train services run in two directions. Accordingly, we extend the L-space representation of the subway infrastructure topology by converting an undirected physical link into a bi-directional flow link, as shown in figures 2(b)-(c).

The weighted adjacency matrix of the passenger flow network $\boldsymbol{F}=\{f_{i\to j}\}$ can be derived from the flow matrix $\boldsymbol{A}$ after introducing the weighted links and directional flows [30-31]. We also introduce the compartment labeled 0 to represent the network exchange with exogenous flows. Note that the exogenous flows, $v_{0\to i}$ and $u_{i\to 0}$, can be derived from flow matrices $\boldsymbol{B}$ and $\boldsymbol{C}$. Accordingly, building a passenger flow network from trip data guarantees that $f_{i\to j}$ always represents passengers taking trips between adjacent stations and $v_{0\to i}$ / $u_{i\to 0}$ always represents passengers entering/leaving one station. Therefore, flow characteristics can be calculated as network properties at the node level.

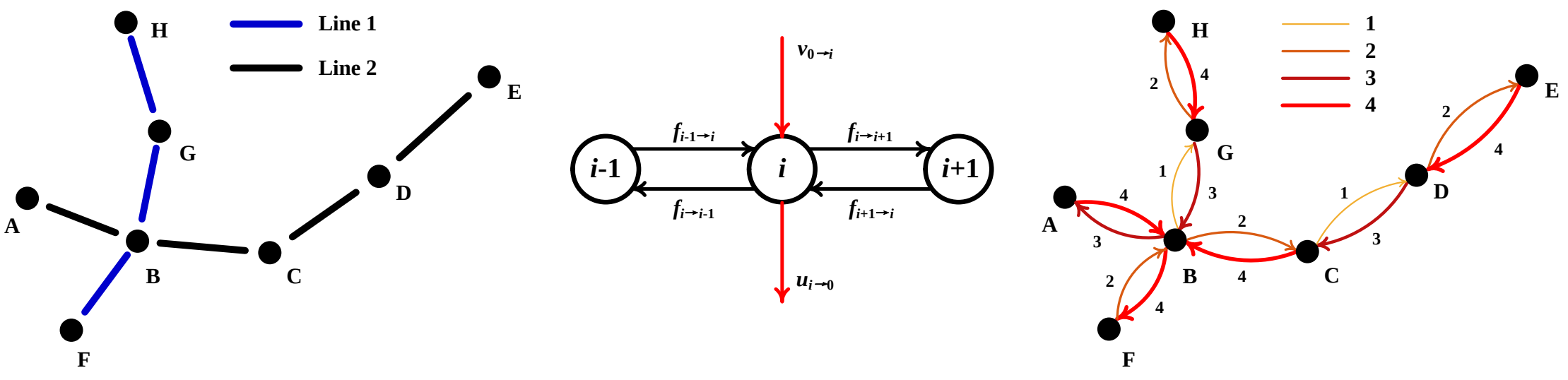

(a) L-space for subway systems (b) Simplified network structure and notation (c) Passenger flow network

Figure 2 (color online) Schematic representations of (a) the subway infrastructure network topology, (b) simplified network structure and notation and (c) the corresponding passenger flow network, with the widths of directed links being proportional to their weights indicated by the corresponding numbers.

## 4. Network characteristics

### *4.1. Flow weight distribution*

The BSS network passenger flow properties are calculated. Figure 3 illustrates the distribution of the number of trips between adjacent subway system stations. The flow weight distribution $P(f_{i \to j})$ is heavy-tailed and displays significant curvature on a log-log scale. The flow weight distribution is fitted by a power-law with an exponent of 1.02. This power-law distribution indicates that the significant passenger movement heterogeneities exist in the subway network, suggesting that individual movements in Beijing represent a heterogeneous flow organization. The similar heterogeneous flow organizations have been observed in the Singapore Mass Rail Transit [24], Seoul Subway [29] and London Underground [27, 31].

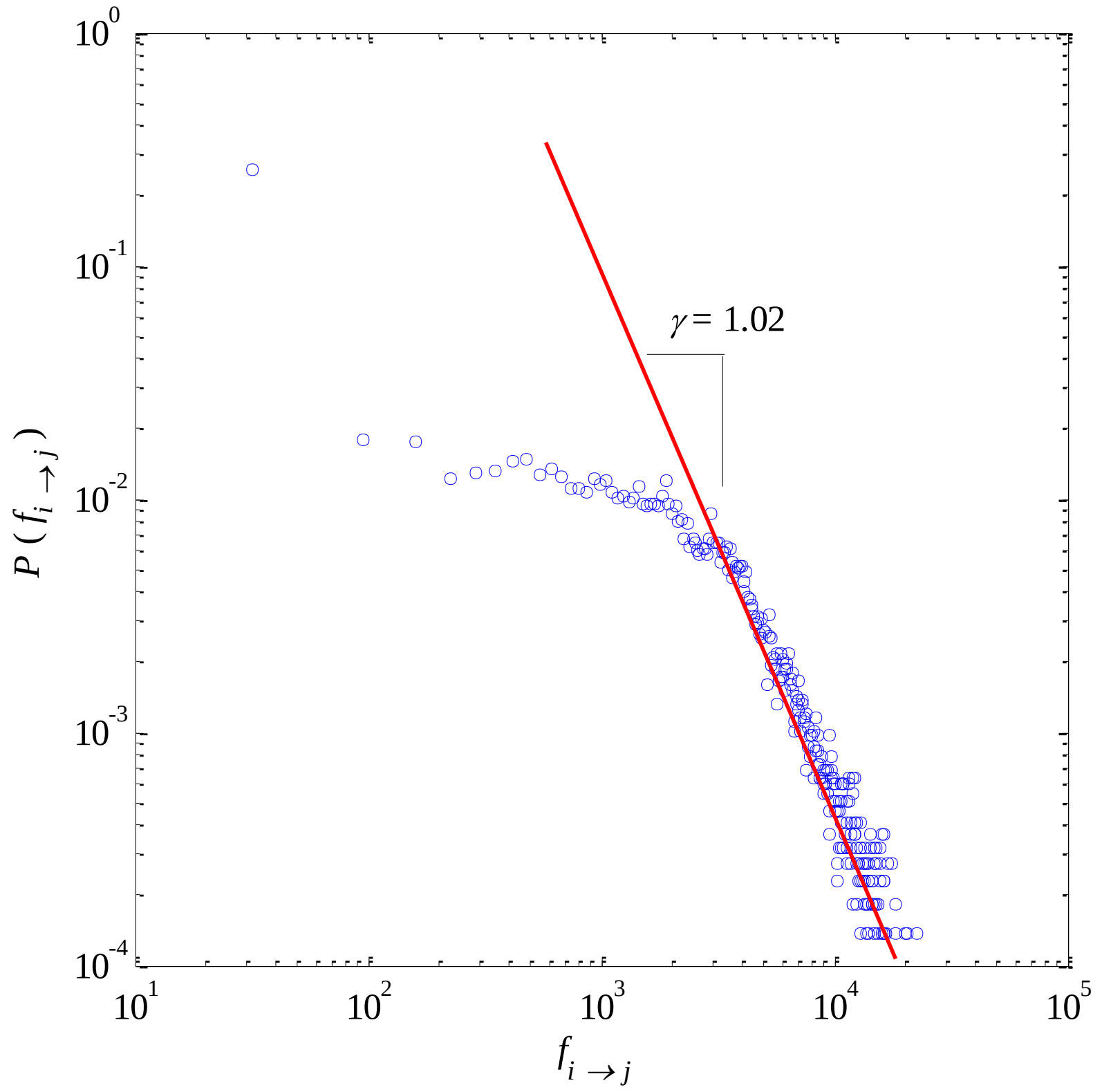

Figure 3 (color online) Flow weight distribution. Log-log plot of the normalized histogram of the number of trips between two adjacent subway stations. The line is fit using the power-law with an exponent of 1.02. The goodness-of-fit ($R^2$) is 0.99.

The incoming and outgoing flow distributions are then investigated for a given station to examine the trip data between two stations. Figure 4 presents the (a) total inflow and (b) total outflow distributions for the subway network. Both the total inflow distribution $P(f_{\cdot \to i})$ and total outflow distribution $P(f_{i \to \cdot})$ exhibit power-law behaviors with exponents of 0.99 and 0.98, respectively. The rank-ordered total time-of-day flows indicate that stations generating large inflows often display larger outflows. The in/outflows are statistically balanced over the entire day, suggesting that most rides are round trips for the commuters. However, the stations often exhibit large inflows and small outflows during the morning peak hours [31]. The exponential distributions of the total flows suggest that individual movements are concentrated in a few centers or hubs, which are dispersed throughout the BSS network as well as the city.

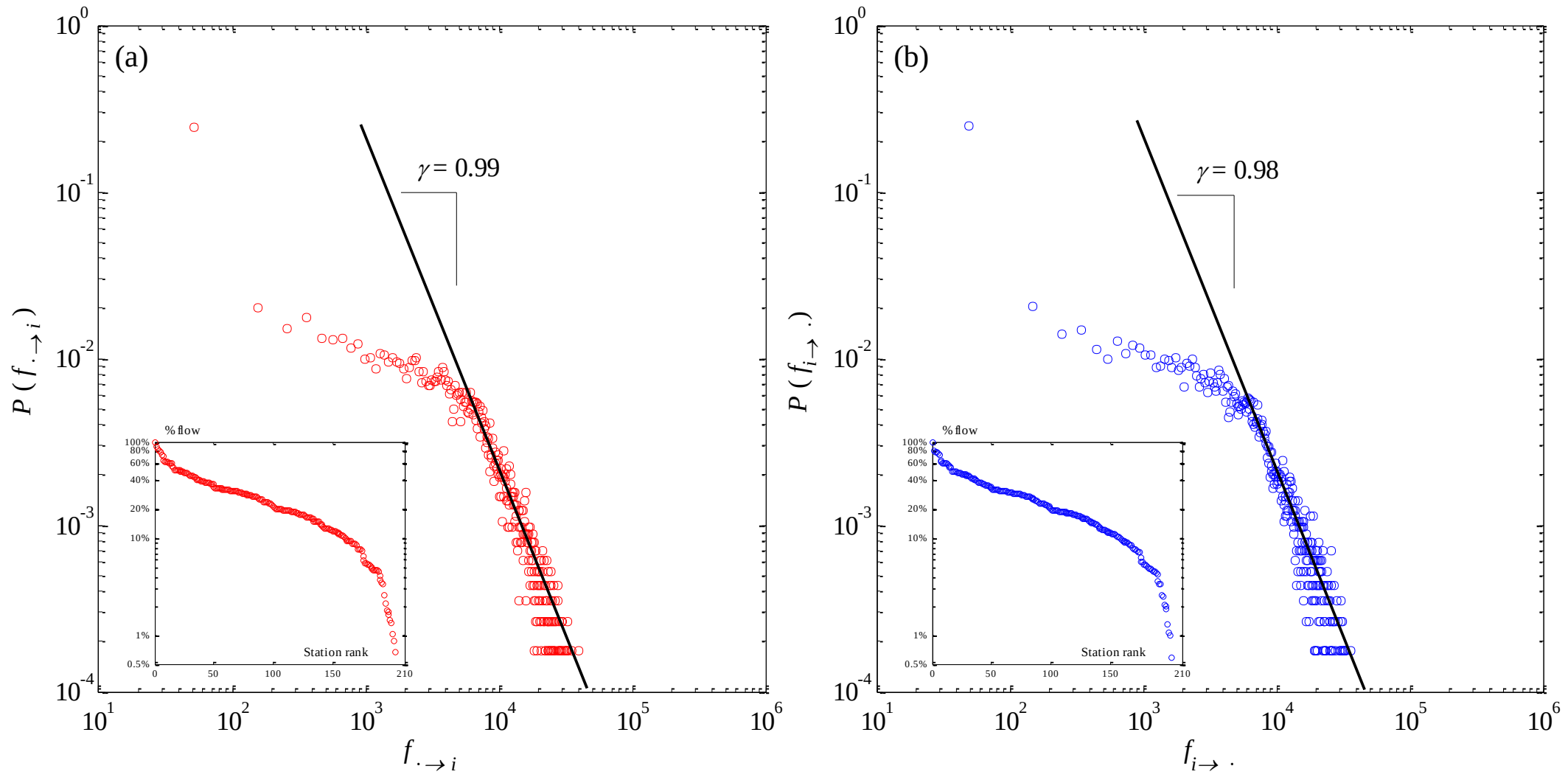

Figure 4 (color online) Total flow distributions. Log-log plots of the (a) total inflows and (b) total outflows. The inflow $f_{\cdot\to i}$ and outflow $f_{i\to\cdot}$ of station $i$ can be defined as $f_{\cdot\to i} = \Sigma_j f_{j\to i}$ and $f_{i\to\cdot} = \Sigma_j f_{i\to j}$, respectively. The lines are visual aids. The lines in (a) and (b) are fit using the power-law with exponents of 0.99 and 0.98, respectively. Both $R^2$ values are 0.99. Insets: Zipf plots of the rank-ordered time-of-day total inflows and total outflows on the lin-log scale.

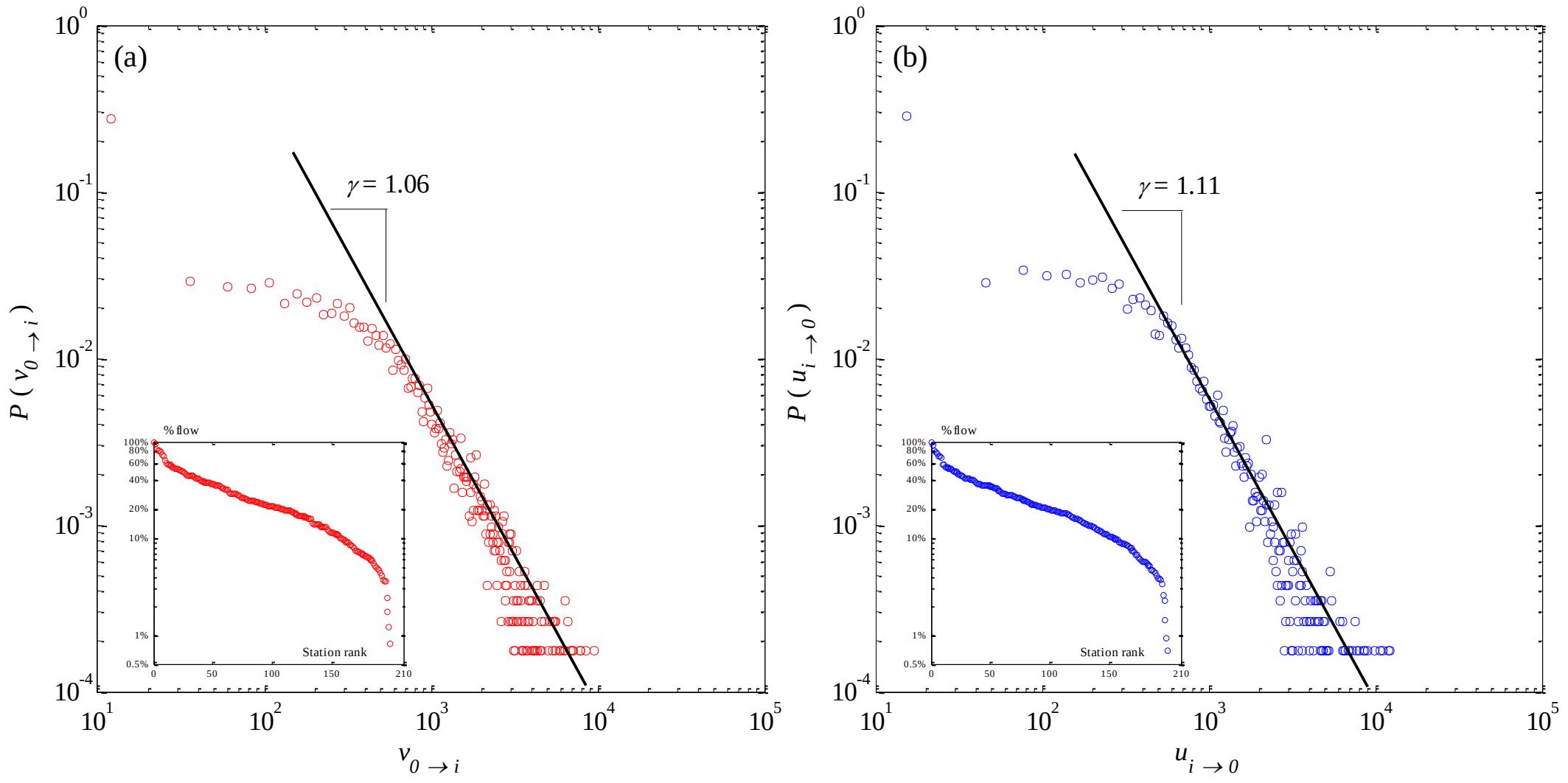

Figure 5 (color online) Exogenous flow distributions. Log-log plots for the (a) flows from outside the network to station $i$ and (b) flows from station $i$ to outside the network. The lines are visual aids. The lines in (a) and (b) are fit using the power-law with exponents of 1.06 and 1.11, respectively. Both $R^2$ values are 0.99. Insets: Zipf plots of the rank-ordered time-of-day exogenous flows on the lin-log scale.

In addition to the trips between two nodes, we also investigate the distributions of the number of passengers entering and leaving a given station. Figure 5 presents the distributions of the (a) flows from outside the network to a station and (b) flows from a station to outside the network. These exogenous flows can be closely approximated by power-law distributions and display an exponential decay on the lin-log scale, which is closely similar to the total flow characteristics. This supports the finding that the network possesses heavy-load centers or hub nodes that experience large external flows. Generally, these stations are located in residential areas or business districts that are easily accessed from other subway lines of BSS network or via other transportation modes such as the bus system.

It is noteworthy that subway passenger movement patterns are strongly heterogeneous in terms of flow size. In fact the BSS network is not a closed system, as it can be considered a subnetwork of a wider transportation network (e.g., subway-bus network). This multimodal urban transit network is scale-free after changing the subway network to take into account the bus system [33]. The topology of the extended transit network is much more heterogeneous than single system of the extended network alone, and some stations are highly concentrated due to the limit of geographic space. This topological “rich-get-richer” phenomenon leads to the fact that there are a few nodes called hub stations in the extended network that experience disproportionately large traffic, and play a dominant role in serving passengers of the overall network. The heterogeneity of the individual movements in the extended transit network will be enhanced when the bus system is coupled with the subway system. In other words, this is equivalent to saying that such enhanced heterogeneity of flow makes the power-law exponent in the extended subway-bus network is smaller than the results we presented in figures 3-6.

### *4.2. Node throughflow distribution*

Node strength is able to represent the node degree in weighted networks. Each node possesses an in-strength and an out-strength, which can be defined as the sum of either the inflows or outflows at a node. The total inflows $f_{\cdot\to i}$ and outflows $f_{i\to\cdot}$ of node $i$ are not equal in a subway network due to capacity constraints. Therefore there are two quantities to keep track of, which we refer to as the *throughflow*, $T_i^{in}$ and $T_i^{out}$ of node $i$, as shown in Equations 4-5.

$T_i^{in}$ and $T_i^{out}$ measure the workload of station $i$ in a subway network based on the sum of passengers moving through station $i$. Figure 6 displays the distributions of (a) $T_i^{in}$ and (b) $T_i^{out}$ for all subway system stations. These distributions are approximately exponential, which is expected because $f_{\cdot\to i}$ ($f_{i\to\cdot}$) and $v_{0\to i}$ ($u_{i\to 0}$) display similar patterns. In particular, comparing with figures 3-6, these flow distributions are statistically self-similar, which indicates that the fractal network of the passenger flows as a whole has the same statistical properties as one or more of the parts. Intuitively, all flow distributions decay according to a power-law that does not change if scales of an observation. This scale-invariance can be explained by the fact that the inherent inhomogeneity and modularity of complex networks is expected to influence the distribution of the flows throughout the network [34-35].

For the BSS network, the modular structure refers to the hierarchical clustering of urban activities and polycentricity, which leads to the highly unevenly distributed passenger person movements in the subway network. We also expect to observe similar results in other subway networks such as Seoul Subway [30] and London underground [31], in biology such as the metabolic network and the protein network [35], and in some technological networks such as Ethernet local area network [36].At larger observation scales, $T_i^{in}$ ($T_i^{out}$) is able to provide an integrated view that includes passengers taking trips between stations and only using a station. This result serves as a basis for hierarchical passenger flows clustering at the next subsection.

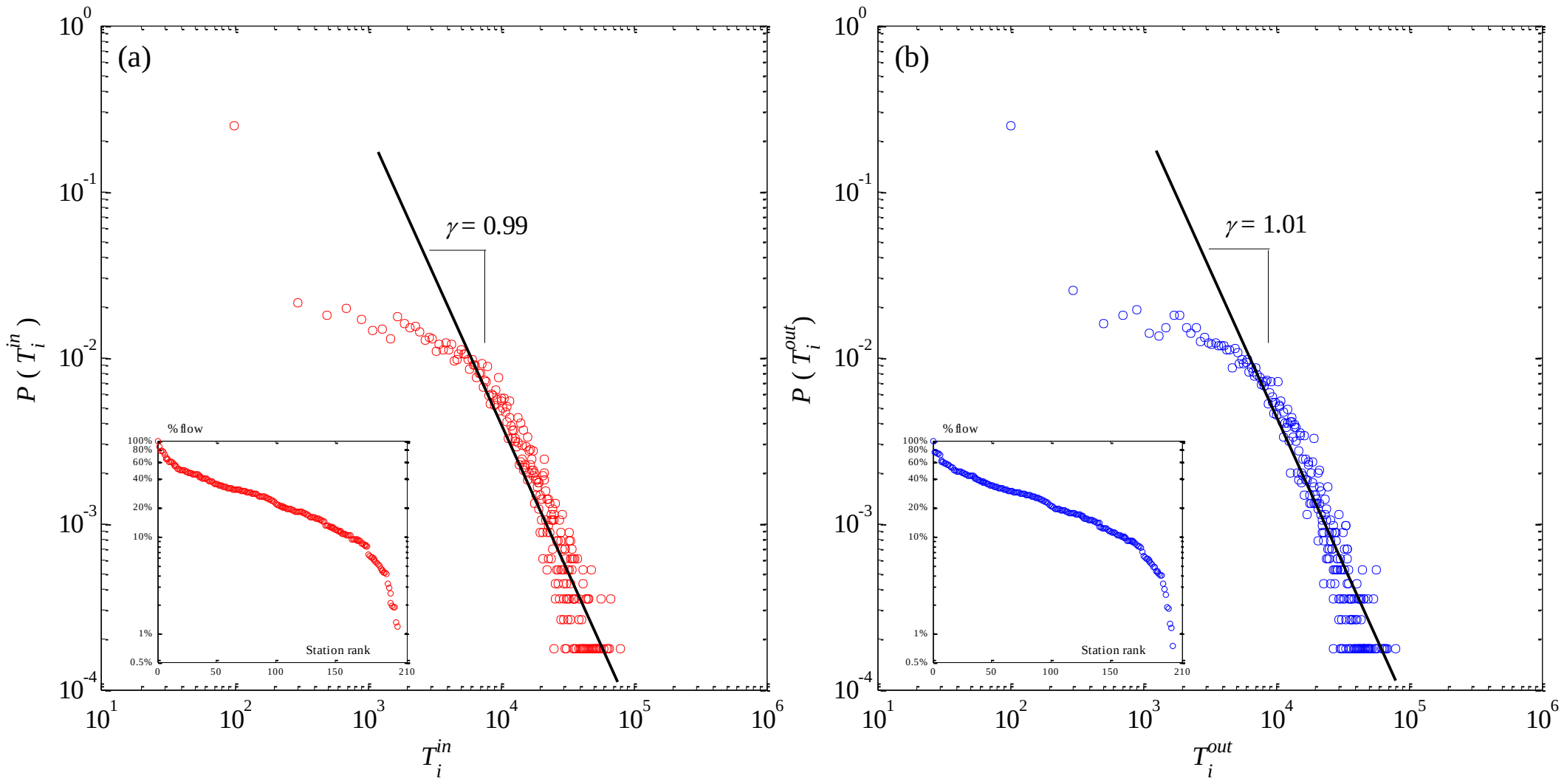

Figure 6 (color online) Node throughflow distributions. Log-log plots for (a) $T_i^{in}$ and (b) $T_i^{out}$ for all nodes. The lines are visual aids. The lines in (a) and (b) are fit using the power-law with exponents of 0.99 and 1.01, respectively. The $R^2$ values are both 0.99. Insets: Zipf plots of the rank-ordered time-of-day throughflow on the lin-log scale.

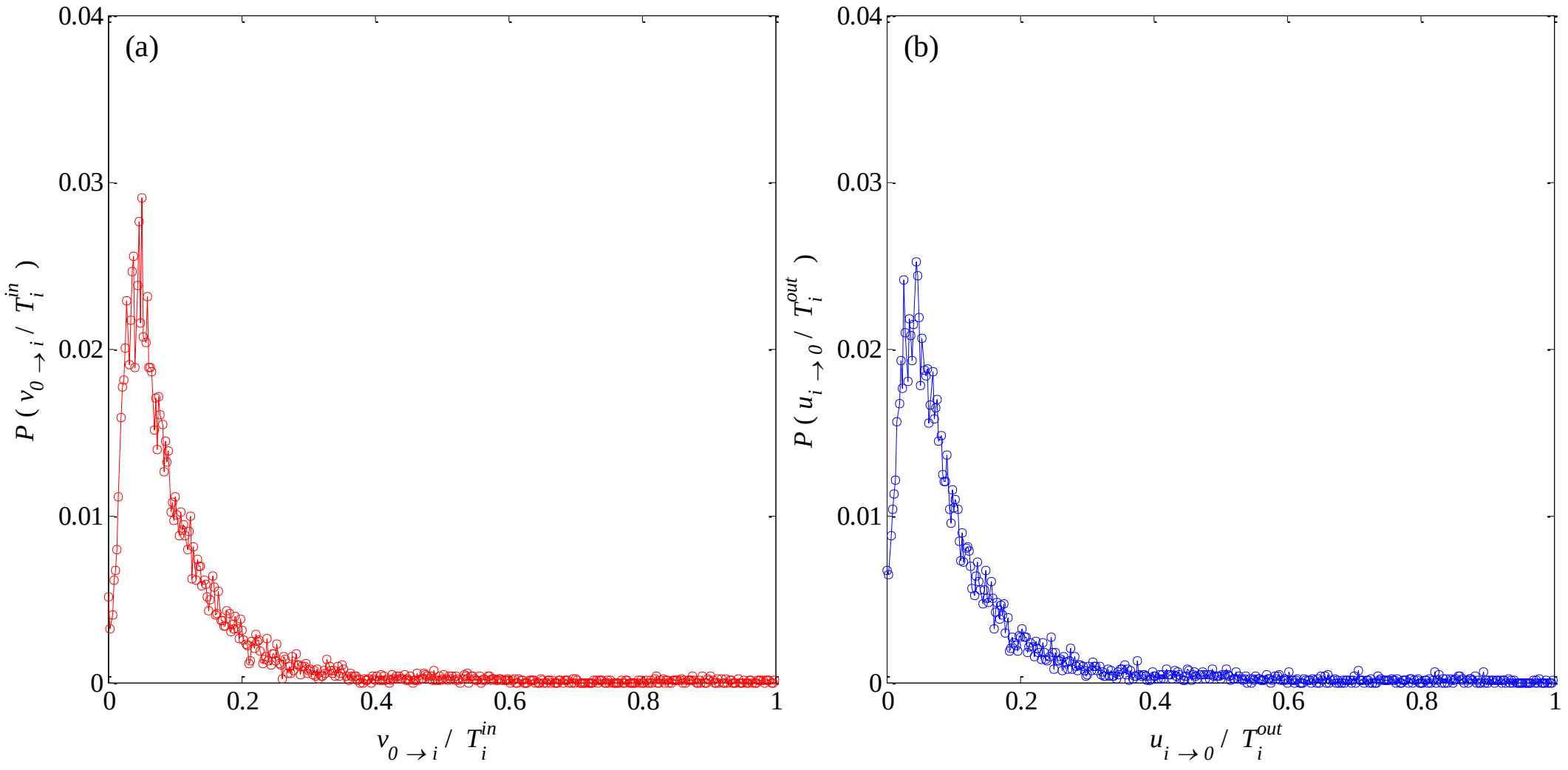

Figure 7 (color online) Probability distribution of (a) $v_{0\to i}/T_i^{in}$ and (b) $u_{i\to 0}/T_i^{out}$. The $P(v_{0\to i}/T_i^{in})$ distribution is fit by a log-logistic distribution with $\alpha$=0.073 and $\beta$=1.878, which corresponds to a mean $\mu$=0.123 and standard deviation $\sigma$=inf. The $P(u_{i\to 0}/T_i^{out})$ distribution is fit using a log-logistic distribution with $\alpha$=0.071 and $\beta$=1.798, which corresponds to $\mu$=0.127 and $\sigma$=inf.

The exogenous flows, $v_{0\to i}$ and $u_{i\to 0}$, between station $i$ and outside the network are very important in the context of platform crowding management or evacuations. These flows are comparable in size to the node throughflow $T_i^{in}$ and $T_i^{out}$. Figure 7 illustrates the (a) $v_{0\to i}/T_i^{in}$ and (b) $u_{i\to 0}/T_i^{out}$ distributions. The first quantity represents the fraction of passengers flowing into station $i$ that entered the same station. And the second quantity represents the fraction of passengers moving out of station $i$ that left the same station. The distribution of $v_{0\to i}/T_i^{in}$ peaks at approximately 0.05 and is fit using a log-logistic distribution. The $u_{i\to 0}/T_i^{out}$ distribution displays a similar behavior. This result suggests that most stations are similar with respect to the fraction of passengers boarding and leaving the station. In addition, most stations exhibit similar passenger capacity utilizations despite significant location, facility, equipment and operational differences.

In many complex networks, the inflow is equal to the outflow at the node level due to flow conservation. However, subway stations cannot efficiently accommodate the passenger demand when the capacity utilization rate is high, causing passenger delays. These delays violate flow conservation principles, namely $T_i^{in} \neq T_i^{out}$, as shown in figure 8(a). The passenger delays occur because: (i) the remaining train capacity is not able to accommodate all waiting passengers and (ii) the departing passengers cannot efficiently leave a station due to in-station facility limitations, such as stairwell or corridor capacities.

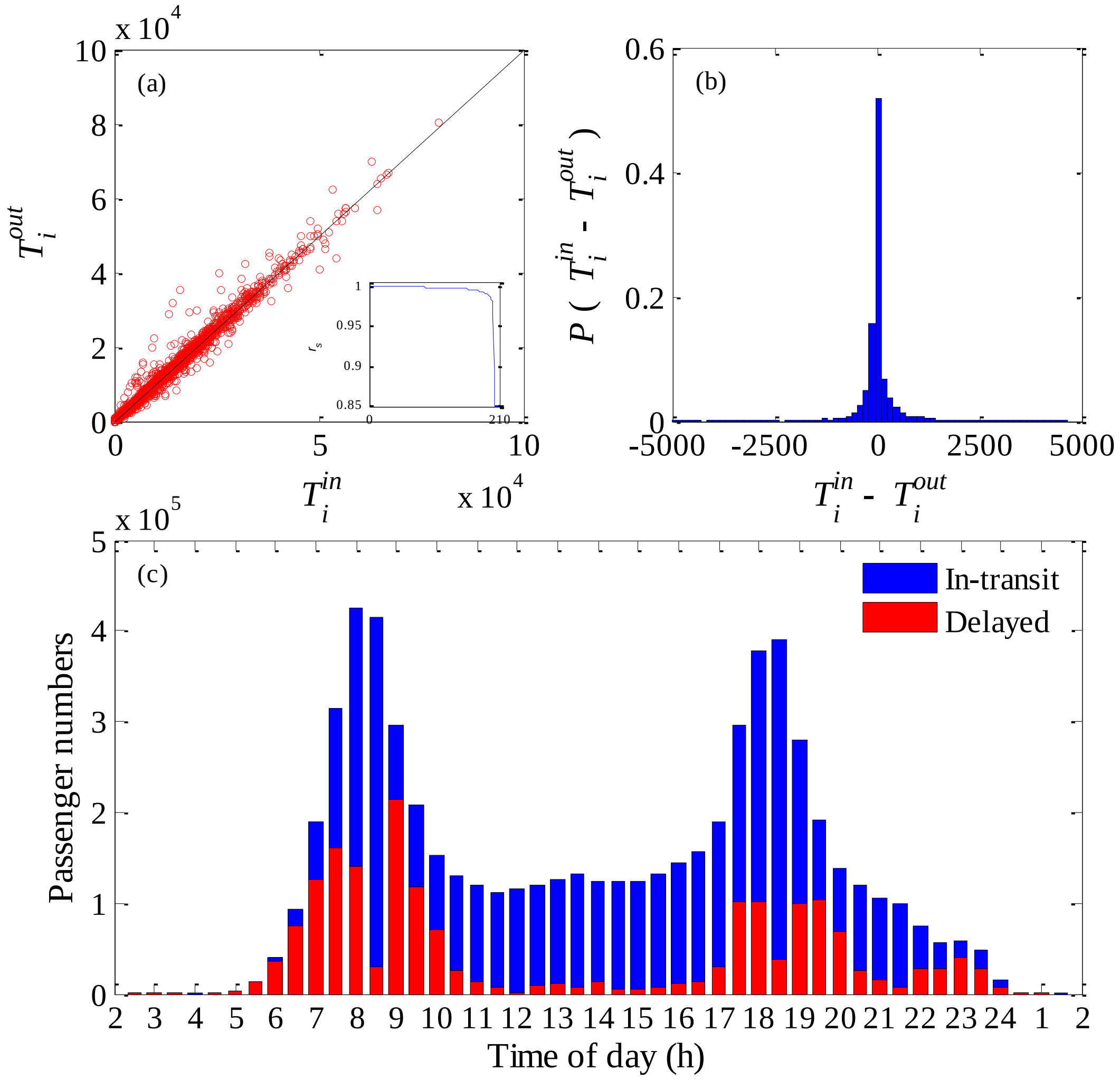

Figure 8 (color online) Temporal patterns of in-transit and delayed passengers. (a) The linear relationship between $T_i^{in}$ and $T_i^{out}$. A strong positive correlation holds for the most studied stations, as more than 90% of the stations display $r_s$>0.95 via a Spearman correlation analysis. (b) Probability distribution of $d_i$ for the entire day. The distribution can be fit using a normal distribution with parameters $\mu$=-6 and $\sigma$=920. (c) In-transit and delayed passenger demands based on time-of-day.

Accordingly, the subway passenger demands can be classified into two groups, which include the delayed passengers and in-transit passengers. The number of delayed passengers can be calculated as the absolute balance between the node throughflows, which is given by $d_i=|T_i^{in}-T_i^{out}|$. The number of in-transit passengers can be estimated as the balance between passengers entering a station and passengers delayed at the station, given by $b_i=v_i - d_i$. Figure 8(b) shows that the distribution of $d_i$ is approximated using a normal distribution, exhibiting a slightly asymmetrical shape and high peakedness.

Figure 8(c) illustrates the total number of in-transit and delayed passengers at all stations based on the time-of-day, which is able to provide a time-volume relationship for both trains and platforms. The total number of in-transit passengers displays a morning peak at 8:00. In addition, an evening peak can be observed at 18:30. The total number of delayed passengers also experience significant morning and evening peaks. This temporal pattern suggests that the demand cannot be effectively and efficiently accommodated during the peak travel hours, causing crowding and delays at central stations or hubs.

These empirical findings provide a basic understanding of BBS network passenger flow characteristics. Furthermore, for any point in time interval and any station, the quantities of in-transit and delayed passengers can be calculated at any time interval and any station, which can be used to develop effective response strategies and improve train service operations.

### *4.3. Hierarchical organization of the passenger flows*

In addition to analyzing the statistics and temporal patterns associated with the flows, we also investigated station clustering. Hierarchical flow organizations are often deemed community structures when studying complex networks. Many methods exist for finding communities in real networks [37], the most common of which are hierarchical clustering methods [38-39]. Hierarchical clustering methods involve similarity measurements between data, which are used to build cluster trees or dendrograms by joining objects, starting with the closest pairs and ending with the farthest. This study uses a hierarchical cluster analysis to group different stations based on the node throughflow $T_i^{in}$. The Euclidean distance acts as the natural similarity measure in our network because stations with passenger movements are located in space. A cluster is created using the agglomerative clustering method with average linkage criterion, which is also called the unweighted pair group method with an arithmetic mean (UPGMA) [40]. Although we produced similar results using other linkage criteria, we only show the results using the UPGMA, which displays the appropriate number of groups.

Figure 9 highlights the hierarchical passenger flow organizations through the subway network and corresponding spatial characteristics of stations in the system The dendrogram in figure 9(a) summarizes the hierarchical clustering results for the morning peak (7:00-10:00). The number of clusters is not an absolute quantity, but depends on an observation scale, which is based on $T_i^{in}$. The hierarchical tree diagram can be clustered into ten groups based on $T_i^{in} \geq 11800$. The following trends can be observed:

(i) The BSS network passenger flows are not evenly distributed. Sixteen of the 203 stations (Groups 1-5) contribute 21% of the flows, whereas the remaining 79% of total flows come from the remaining 187 stations (Groups 6-10). For example, Group 1 contains only *Guomao* station, which is the most heavily burdened station in the network at more than 1.9% of total flows. The fractions of total flows produced from the Groups 2-10 are 4.6%, 6.4%, 5.6%, 2.0%, 23.1%, 13.6%, 17.3%, 16.8% and 8.7%, respectively.

(ii) Groups 1-5 appear form a hierarchy in which the five communities are members of a larger heavy-burden stations community. As displayed in figure 9(b), the 16 stations are all transfer stations, which are located in either residential areas or business districts in Beijing. High transfer passenger demands during the morning peak lead to large concentrations of the commuting passenger at these center stations or hub nodes, which are close to major business buildings and public transportation centers.

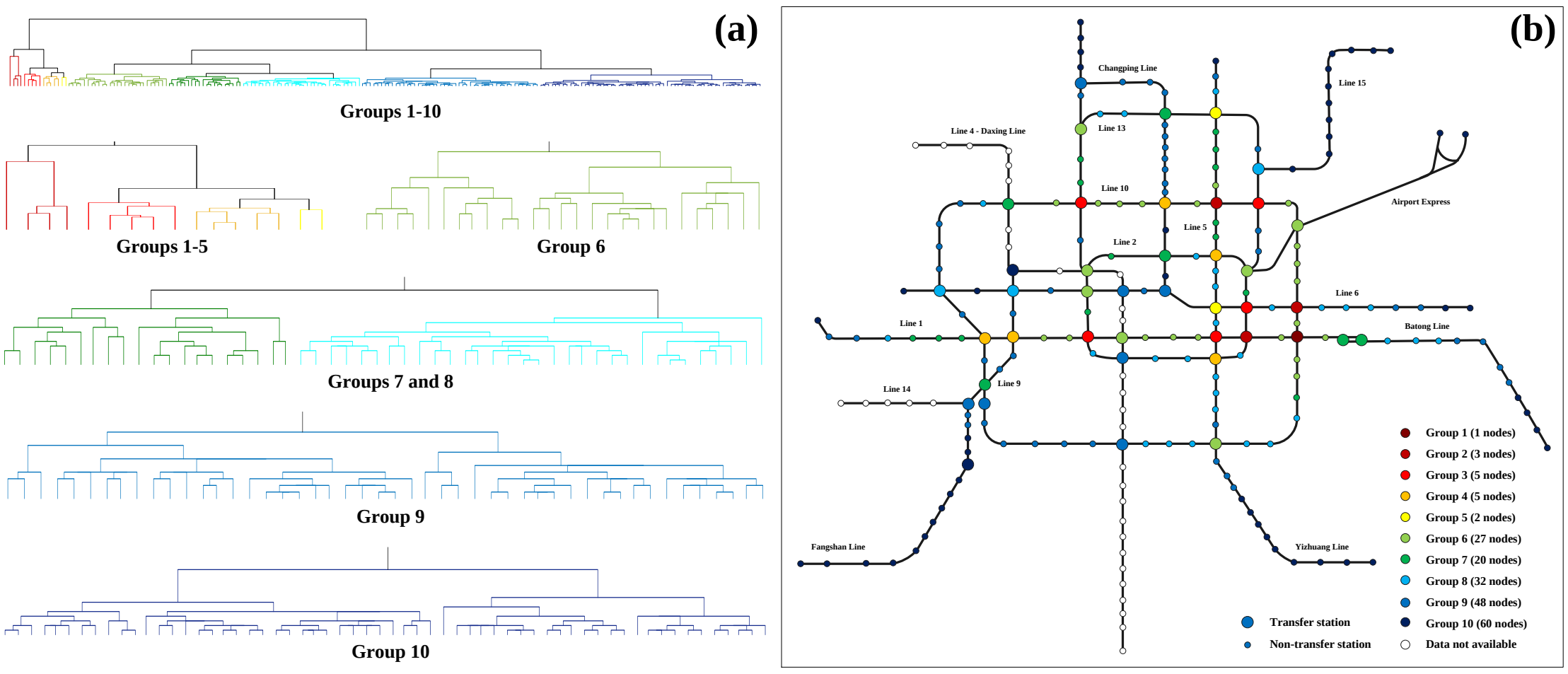

Figure 9 (color online) Hierarchical organizations of the passenger flows during the morning peaks. The ten clustering groups defined in the dendrograms (a) for the morning peak and (b) the corresponding spatial distribution of the passenger flow organization. The vertical axis of the dendrogram measures the node throughflow $T_i^{in}$. $T_i^{in}\geq 11800$ was set as the clustering scale for the morning peak.

(iii) Groups 6-8 form a larger clustering group. Most stations on Lines 1, 2, 5 and 10 can be categorized into this clustering group. Note that Line 1 runs beneath the busiest east-west thoroughfare of the city. Line 5 is a straight north-south line that can relieve the north-south traffic pressure. Lines 2 and 10 are both rectangular loop lines around the 2$^{nd}$ and 3$^{rd}$ Ring Roads. Stations on these lines transport large quantities of passengers.

(iv) Two large communities, Groups 9-10, occur within the BSS system branch lines. The Beijing population is largely distributed in various suburbs and residential areas. As the system grows, branch lines radiate from the city center and reach new urbanized areas. The commuting passenger flow patterns between domestic suburban areas and urban workplaces exhibit similar properties throughout the city, which are dominated by two-way asymmetric morning and evening peaks flows into and out of the center of the Beijing metropolitan area.

The BSS network spatial passenger flow distributions that correspond to the hierarchical throughflow organization are shown in figure 9(b) for morning peak hours. The passenger flow sizes drastically differ in the spatial domain. The majority of the morning peak demand originates from branch line stations (Groups 9-10) that originate in suburban areas. The majority of the passengers travel to stations located in central urban areas (Groups 6-8), especially the stations located in the central business districts or hub stations (Groups 1-5). The evening peak hour (17:00-20:00) passenger flows also reveal a hierarchical structure, which indicates that the demand originates from workplaces in central urban areas and diffuses into domestic suburban areas, as shown in figures 10(c) and (d). These two figures suggest that the most visited stations represent daily commuting options that passengers use to travel to work in the morning and return in the evening.

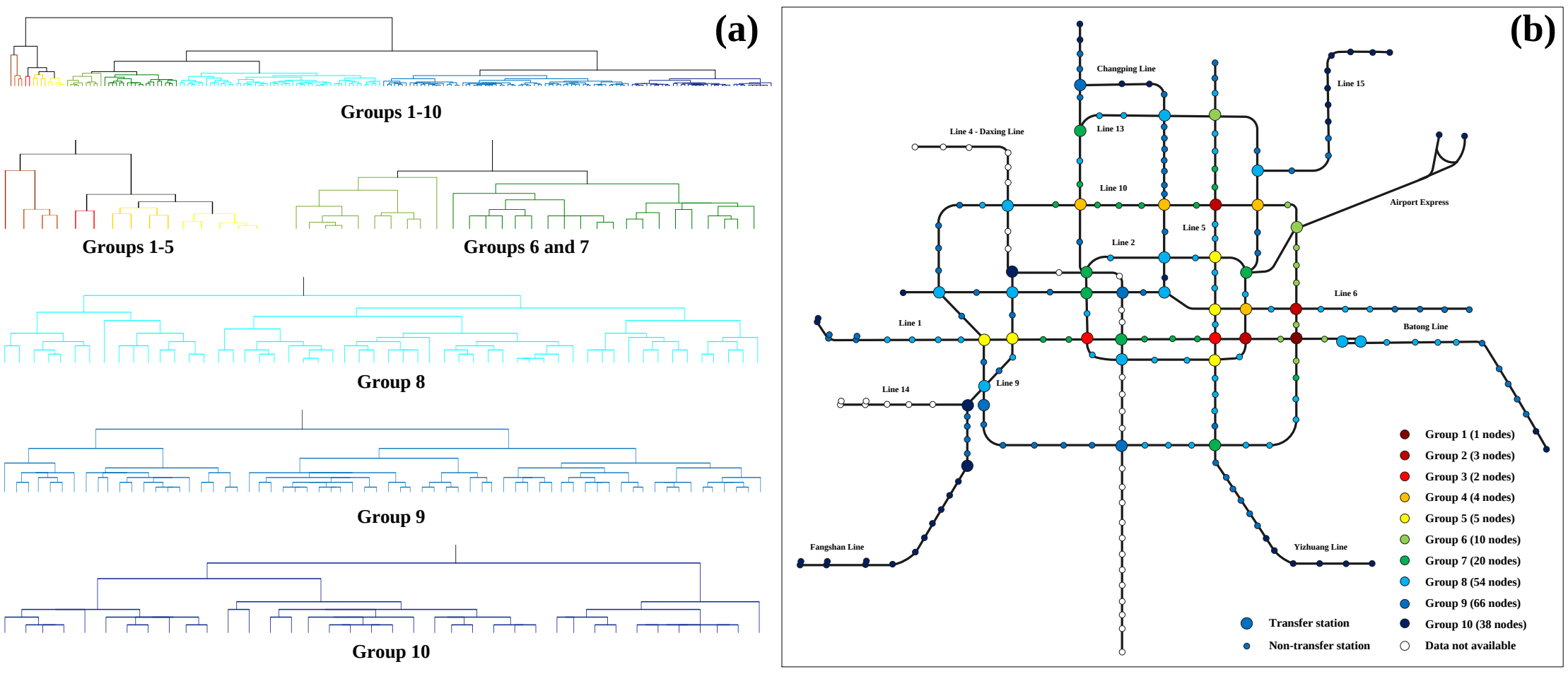

Figure 10 (color online) Hierarchical organizations of the passenger flows during the evening peaks. The ten clustering groups defined in the dendrograms (a) for the evening peak and (b) the corresponding spatial distribution of the passenger flow organization. The vertical axis of the dendrogram measures the node throughflow $T_i^{in}$. $T_i^{in}\geq 10500$ was set as the clustering scale for the evening peak.

This hierarchical cluster analysis demonstrates that most of the morning peak trips are associated with the residential and suburban areas. Conversely, the majority of the evening peak trips are associated with downtown or business district stations. During a typical weekday, the traffic flows in and out of the city center are due to commuters who travel to work during the morning peak and return home during the evening peak. These spatial mobility patterns correlate with land use in the Beijing metropolitan region, indicating the urban polycentricity hierarchy. Therefore, the ten most important polycenters in Beijing are examined using the simple clustering algorithm described in [31], as shown in figure 10. It is noted that this algorithm can be viewed as a special case of a scale-invariant renormalization procedure [34, 41] when taking into account the geographical proximity of groups of stations and a scale of observation measured by the fraction of total node throughflow.

Per figure 11, the most visited places are distributed along Chang'an Avenue, the large east-west thoroughfare that transects the city center. These places include *Jianguomen*, *Gongzhufen*, *Guomao*, *CCTV Tower* and *Beijing Financial Street*, which can account for 36.7% of the total throughflow. Geographically situated to the east of the city center, the Beijing Central Business District (Beijing CBD) currently accounts for 25.3% of the total throughflow. *Zhongguancun*, the China's Silicon Valley, in northwest Beijing is another highly visited place. The remaining four polycenters are related to the predominant residential neighborhoods to the north of the city. These ten most visited places account for as much as 60% of the total throughflow. In general these findings confirm that the Beijing metropolitan area exhibits a polycentric urban structure, which agrees with previous empirical findings [42]. This trend is also supported by city's population redistribution [43-44].

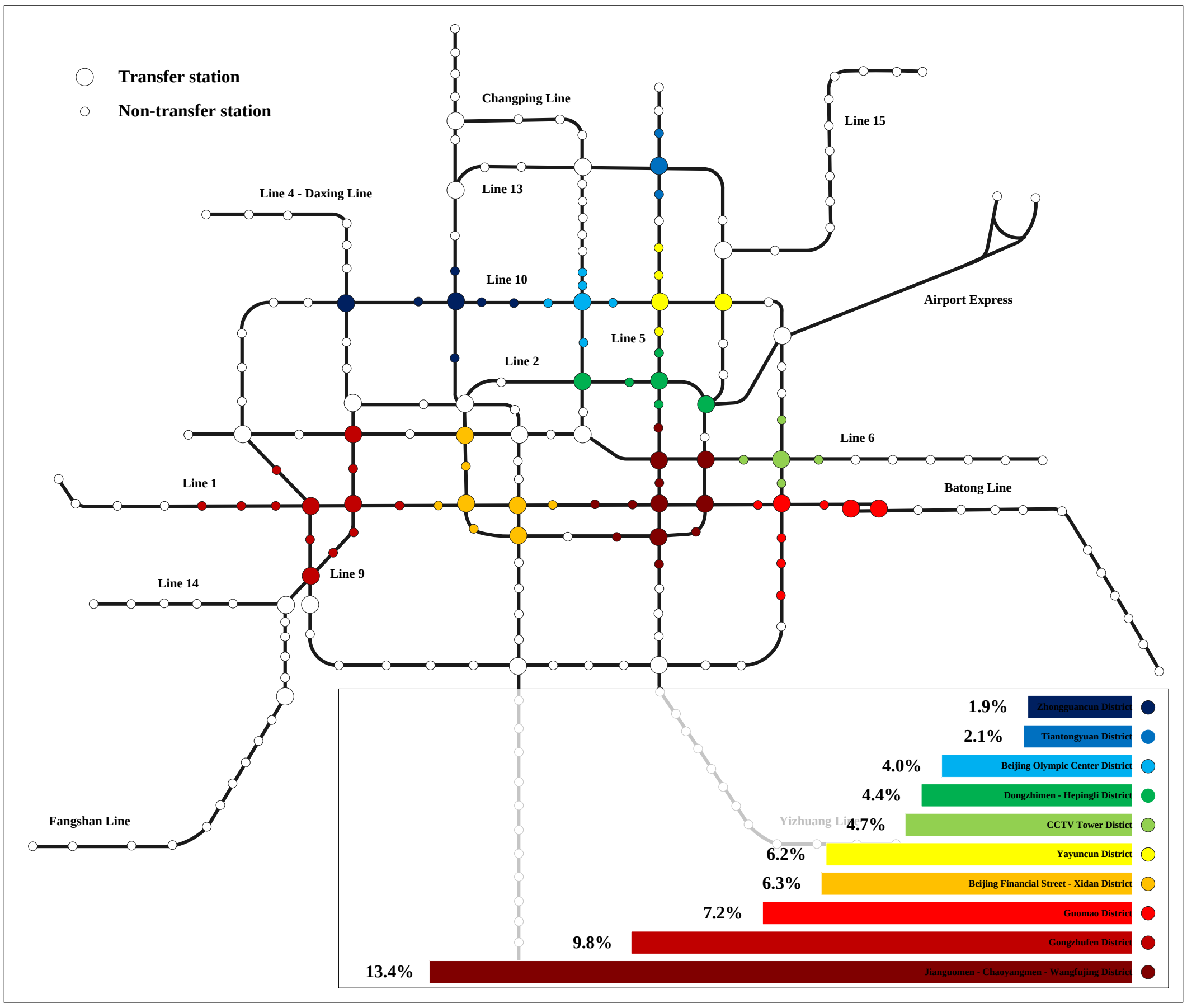

Figure 11 (color online) The Beijing Subway System: Hierarchical organization of urban activities and polycenters. The ten most important polycenters in Beijing are illustrated by ranking stations in descending order based on node throughflow, $T_i^{in}$, and all stations within 2 km of a pre-defined center are aggregated (ten most high throughflow stations), representing 60% of the total node throughflow. Inset: the fraction of the total node throughflow for these polycenters.

The hierarchy of urban polycenters highlights a hierarchical, descending decomposition of the node throughflow with respect to an increasing share of the total node throughflow in the BSS network as well as the city. In fact, the number of the urban centers is not an absolute quantity, but depends on an observation scale. At large spatial scales, we are able to observe a large center corresponding to the whole city, and when we decrease the scale of observation, multiple centers of urban activities would emerge, which are themselves composed of smaller centers. For example, we can find ten most important polycenters over the whole region of the Beijing city, three multiple centers in the Beijing CBD, situated to east of the Beijing, and two multiple centers in *Guomao* district, situated to east of the Beijing CBD.

The hierarchical nature of the network of passenger flows is crucial and indicates that the centers of urban activities can be found at many scales of observations. This scale-invariance of the BSS network is closely similar to self-similar character of the fractal biology networks [35, 41] or the non-fractal technological networks [34] under the renormalization scheme. It indicates that transportation systems share common features of other natural and man-made complex networks, which can be understood by applying the renormalization transformation that enable one to exploit the self-similarity of the fractal transportation networks.

The spatial distribution of urban centers is an important factor when analyzing individual movements within Beijing. The inner-city redevelopment and suburbanization in emerging megacities such as Beijing demonstrate that political, commercial, cultural and other core functions are centralized in the inner cities, while residential areas are located along the outskirts of the cities. The concentration of urban activities leads to highly heterogeneous spatial mobility patterns and flow sizes. These heterogeneous flow patterns have also been observed in Singapore, Seoul and London [24, 27, 29-31], indicating that commuters flow into and out of city centers during the morning and evening peaks. Based on the results of this analysis, we conclude that the commuting patterns, such as individual flow entrances and exits, represent a hierarchical urban polycentricity.

In emerging metropolises such as Beijing, urban public transportation is thriving as the communities grow. Using the Yikatong data we can characterize the heterogeneous passenger flow organizations, then identify hierarchical organization of urban polycenters in the Beijing city as a ranking of different places (i.e., the clustering groups of subway stations) based on the number of passengers flowing into these centers. General observation suggests that these complexities of traffic flow might apply to other world metropolises. Our analysis needs to be extended to include travel information of other transportation modes such as the bus system, which will provide a complete view of the polycentric activities and entangled hierarchical flows within Beijing, and then enrich polycentricity.

## 5. Summary and conclusions

Network methods are useful for studying transportation systems. This study applied a network methodology to individual passenger flows within a large subway network. A directed weighted network of passenger movement was created for the Beijing Subway System using a trip dataset collected from smart subway card transactions. The flow weight and node throughflow distributions were characterized to identify the temporal patterns of in-transit and delayed passengers, which is able to serve as a reference for researchers and engineers. The hierarchical flow organizations were also examined, which identified groups based on the number of passengers flowing into the stations, as well as the spatial flow distributions. The corresponding spatial mobility patterns represent a hierarchical urban polycentricity, which is common in megacities.

The empirical findings provide insights regarding urban human mobility patterns within a large subway network. Our approach provides a general framework for studying individual movement patterns in public transportation systems (bus, tram, trolley, etc). Our methodology can be applied to these systems by simply replacing the weights based on other data resources. However, our approach needs to be expanded to incorporate other transportation modes, and to direct the interest to the multiplex characteristics of the extended multimodal transportation systems such as the dynamic robustness, which will enrich the analysis and include additional traffic flow complexities within Beijing.

## Acknowledgments

We acknowledge the support of this paper by the National Basic Research Program of China (No.2012CB725406), the National Natural Science Foundation of China (No.71131001). We also would like to thank the anonymous reviewers for their valuable comments and suggestions that lead to a substantially improved manuscript.

## References

[1] Rosvall M, Trusina A, Minnhagen P and Sneppen K, 2005 *Phys. Rev. Lett*. **94** 028701
[2] Gastner M T and Newman M E J, 2006 *Eur. Phys. J. B* **49** 247
[3] Sen P, Dasgupta S, Chatterjee A, Sreeram P A, Mukherjee G and Manna S S, 2003 *Phys. Rev. E* **67** 036106
[4] Li W and Cai X, 2007 *Physica A* **382** 693
[5] Amaral L A N, Scala A, Barthelemy M and Stanley H E, 2000 *P. Natl. Acad. Sci. USA* **97** 11149
[6] Lin J and Ban Y, 2014 *Physica A* **410** 302
[7] Sienkiewicz Jand Hołyst J A, 2005 *Phys. Rev. E* **72** 046127
[8] Von Ferber C, Holovatch T, Holovatch Y and Palchykov V, 2009 *Eur. Phys. J. B* **68** 261
[9] Derrible S, 2012 *PloS ONE* **7** e40575
[10] Wu J, Gao Z, Sun H and Huang H, 2004 *Mod. Phys. Lett. B*, **18** 1043
[11] Louf R, Roth C and Barthelemy M, 2012 *PloS ONE* **7** e102007
[12] Latora V and Marchiori M, 2002 *Physica A* **314** 109
[13] Derrible S and Kennedy C, 2010 *Physica A* **389** 3678
[14] Latora V and Marchiori M, 2001 *Phys. Rev. Lett.* **87** 198701
[15] Vragović I, Louis E and Díaz-Guilera A, 2005 *Phys. Rev. E* **71** 036122
[16] Ek B, VerSchneider C and Narayan D A, 2013 *Physica A* **392** 5481
[17] Latora V and Marchiori M, 2005 *Phys. Rev. E* **71** 015103
[18] Yang Y, Liu Y, Zhou M, Li F and Sun C, 2015 *Safety Sci.* **79** 149
[19] Leng B, Zhao X and Xiong Z, 2014 *Europhys. Lett.* **105** 58004
[20] Kurant M and Thiran P, 2006 *Phys. Rev. Lett.* **96** 138701
[21] Barthélemy M, 2011 *Phys. Rep.* **499** 1
[22] Lin J and Ban Y, 2013 *Transport Rev.* **33** 658
[23] Kurant M and Thiran P, 2006 *Phys. Rev. E* **74** 036114
[24] Soh H, Lim S, Zhang T Y, Fu X J, Lee G K K, Hung T G G, Di P, Prakasam S and Wong L S, 2010 *Physica A* **389** 5852
[25] Gonzalez M C, Hidalgo C A and Barabasi A L, 2008 *Nature* **453** 779
[26] Alexander L, Jiang S, Murga M and González M C, 2015 *Transport. Res. C* **58** 240
[27] Hasan S, Schneider C M, Ukkusuri S V and González M C, 2013 *J Stat. Phys.* **151** 304
[28] Sun L, Lee D H, Erath A and Huang X, 2012 *Proc. ACM SIGKDD Int. Workshop Urban Computing* (Beijing) (New York: Association for computing Machinery) p 142
[29] Lee K, Jung W S, Park J S and Choi M Y, 2008 *Physica A* **387** 6231
[30] Lee K, Goh S, Park J S, Jung W S and Choi M Y, 2011 *J. Phys. A* **44** 115007
[31] Roth C, Kang S M, Batty M and Barthélemy M, 2011 *PloS ONE* **6** e15923
[32] Fath B D, Patten B C and Choi J S, 2001 *J. Theor. Biol*. **208** 493
[33] Huang A, Zhang H M, Guan W, Yang Y and Zong G, 2015 *Math. Probl. Engineer.* **2015** 940795
[34] Gallos L K, Song C and Makse H A, 2007 *Physica A* **386** 686
[35] Gallos L K, Song C, Havlin S and Makse H A, 2007 *P. Natl. Acad. Sci.* **104** 7746
[36] Leland W E, Taqqu M S, Willinger W and Wilson D V, 1994 *IEEE ACM T. Network.* **2** 1
[37] Fortunato S, 2010 *Phys. Rep.* **486** 75
[38] Sales-Pardo M, Guimera R, Moreira A and Amaral L A N, 2007 *P. Natl. Acad. Sci. USA* **104** 15224
[39] Friedman J, Hastie T and Tibshirani R, 2001 *The elements of statistical learning* (Berlin: Springer)
[40] Sokal R and Michener C, 1958 *University of Kansas Science Bulletin* **38** 1409
[41] Song C, Havlin S and Makse H A, 2005 *Nature* 433 392
[42] Qin B and Han S S, 2013 *Urban Stud.* **50** 2006
[43] Wang F and Zhou Y, 1999 *Urban Stud.* **36** 271
[44] Feng J 2003 *The internal urban spatial restricting in China* PhD thesis, Peking University [in Chinese]